\documentclass[aps,prc,twocolumn,showpacs,floatfix,superscriptaddress,nofootinbib,longbibliography]{revtex4-1}
\usepackage[utf8]{inputenc}

\usepackage{amsmath}
\usepackage{amssymb}
\usepackage{inputenc}
\usepackage{amssymb}
\usepackage{graphicx}
\usepackage{epsfig}
\usepackage{bm}
\usepackage{color}
\usepackage{float}
\usepackage{dcolumn}
\usepackage{multirow} 
\usepackage{MnSymbol}
\usepackage[unicode=true,
  linktocpage,
  linkbordercolor={0.5 0.5 1},
  citebordercolor={0.5 1 0.5},
  colorlinks=true,
  linkcolor=blue,
  citecolor=blue,
  urlcolor=blue]{hyperref} 
\usepackage{lipsum} 
\usepackage{placeins} 
\usepackage[dvipsnames,usenames]{xcolor}

\newcommand{\be}{\begin{equation}}
\newcommand{\ee}{\end{equation}}
\newcommand{\ba}{\begin{eqnarray}}
\newcommand{\ea}{\end{eqnarray}}

\begin{document}
\title{Spectral function for $^4$He using the Chebyshev expansion in coupled-cluster theory}
\date{\today}
\author{J.~E.~Sobczyk}
\affiliation{Institut f\"ur Kernphysik and PRISMA$^+$ Cluster of Excellence, Johannes Gutenberg-Universit\"at, 55128
  Mainz, Germany}
  
\author{S.~Bacca}
\affiliation{Institut f\"ur Kernphysik and PRISMA$^+$ Cluster of Excellence, Johannes Gutenberg-Universit\"at, 55128
  Mainz, Germany}
  \affiliation{Helmholtz-Institut Mainz, Johannes Gutenberg-Universit\"at Mainz, D-55099 Mainz, Germany}
  
 \author{G. Hagen} 
\affiliation{Physics Division, Oak Ridge National Laboratory,
Oak Ridge, TN 37831, USA} 
\affiliation{Department of Physics and Astronomy, University of Tennessee,
Knoxville, TN 37996, USA} 

\author{T.~Papenbrock}
\affiliation{Department of Physics and Astronomy, University of Tennessee,
Knoxville, TN 37996, USA} 
\affiliation{Physics Division, Oak Ridge National Laboratory,
Oak Ridge, TN 37831, USA}

\begin{abstract}

We compute spectral function for $^4$He by combining coupled-cluster theory with an expansion of integral transforms into Chebyshev polynomials. Our method allows to estimate the uncertainty of spectral reconstruction.
The properties of the Chebyshev polynomials make the procedure numerically stable and considerably lower in memory usage than the typically employed Lanczos algorithm.
We benchmark our predictions with other calculations in the literature and with electron scattering data in the quasi-elastic peak.
The spectral function formalism allows one to extend ab-initio lepton-nucleus cross sections into the relativistic regime. This makes it a promising tool for modeling this process at higher energy transfers. 
The results we present open the door for studies of heavier nuclei, important for the neutrino oscillation programs.

\end{abstract}

\maketitle

\section{Introduction}
Lepton-nucleus cross sections are not only a invaluable tool to investigate the nuclear dynamics with clean electroweak probes~\cite{BaccaPastore}, but  have also become a hot topic in the short- and long-baseline neutrino programs aiming at extracting neutrino oscillation parameters~\cite{Nustec,whitepaper1,whitepaper2}.
Recently, we initiated a theory program that addresses low- and intermediate- energy lepton-nucleus scattering from first principles by combining the Lorentz-integral-transform  with coupled-cluster theory  (LIT-CC) to compute many-body response functions~\cite{bacca2013} and lepton-nucleus cross sections.
In this approach, 
 the final state interaction  is described  consistently with the initial state interaction by the same Hamiltonian rooted into quantum chromodynamics, see Refs.~\cite{Acharya:2019fij,Sobczyk:2020qtw,Sobczyk:2021dwm}. An example is the description of the longitudinal quasi-elastic peak of $^{40}$Ca, see Ref.~\cite{Sobczyk:2021dwm},
 and this paves the way for further investigations.
Despite its success,  this approach also has its limitations: the formalism is based on the non-relativistic theory and at the moment is capable of predicting only inclusive cross-sections. Efforts to include the spectrum of the outgoing nucleons below and above pion production are still lacking in ab-initio calculations.
Several other approximations and phenomenological methods instead offer a way to answer such questions~\cite{Nieves:2011pp, Martinez:2005xe,Amaro:2004bs,Leitner:2008ue, Benhar:2005dj}, chief among them being the spectral functions formalism. While spectral functions can be computed phenomenologically~\cite{Benhar:1994hw,Nieves:2017lij, Buss:2007ar}, calculations were performed recently within the self-consistent Green function (SCGF) method~\cite{Rocco:2018vbf,Barbieri:2019ual} using a similar Hamiltonian as in Ref.~\cite{Sobczyk:2021dwm} for the initial state. In the past also the LIT method combined with hyperspherical harmonics was used to obtain the proton spectral function of $^4$He~\cite{Efros:1998eb}.

The main advantage of the spectral function formalism lies in the possibility of detaching the high-energy physics from the ground-state properties of the nucleus, under the assumption that the final state interactions can be neglected. This not only allows one to make predictions for the quasi-elastic peak using the relativistic kinematics and currents, but this approach can also be used at higher energies, e.g., above  the pion production threshold.  Hence, developing efficient ab-initio methods to compute spectral functions deserves attention, which goes beyond the mere fact that their calculation in a many-body system is  {\it per se}  an interesting and challenging task. In this work, we present an approach to the computation of spectral functions which 
opens up the possibility of using ab-initio many-body methods in the  high-energy regime.

The reconstruction of nuclear response functions requires information about the excited states of the system, but usually these are not easily accessible. To circumvent this issue, the problem has been often reformulated  by computing integral transforms of the response function, with Lorentz and Laplace kernels being popular choices~\cite{Efros:1994iq,efros2007,Carlson_1992,Carlson2015,Lovato_2016}. The computation of the integral transform requires one to (only) solve a bound-state problem. The inversion of the transform, needed to obtain the response function, has to be  performed numerically. While accurate results  have been obtained for a variety of electroweak observables~\cite{bacca2013,bacca2014,Lovato_2016,Lovato_2020,Sobczyk:2021dwm},
 the inversion introduces an additional numerical error and is  most stable when the response function exhibits only one or two broad peaks~\cite{bacca2002,bacca2004}. This scenario was recently explored using machine-learning techniques~\cite{raghavan2021}.
Spectral functions often have a more complicated structure and this makes  the inversion of the integral transform difficult.
Therefore, we propose here to use a different
approach that is based on the Chebyshev expansion of the integral kernel (ChEK) introduced in Refs.~\cite{Roggero:2020qoz,Sobczyk:2021ejs}. Although it relies on the idea of the integral transform, it does not require its inversion. Moreover, for a given desired resolution of the spectral reconstruction it allows one to estimate an uncertainty. 

The information about the (discrete) spectrum of excited states in a many-body system can be retrieved in various ways.
The nuclear theory community is familiar with the Lanczos orthogonalization procedure~\cite{Lanczos:1950zz}, which for example is used in the LIT-CC method. An alternative 
approach, developed in the field of condensed-matter physics,  is the kernel polynomial method (KPM)~\cite{Wei_e_2006}. 
As for the KPM, the approach of this paper is also based on an expansion in Chebyshev polynomials.

This paper is organized as follows. In Sec.~\ref{sec:lepton-nuc_scattering}, we review how the lepton-nucleus interaction in the quasi-elastic peak can be expressed in terms of spectral functions using the  impulse approximation. In Sec.~\ref{sec:formalism}, we present the theoretical framework for our calculations of spectral functions. We validate our method in $^4$He, paying special attention to the center-of-mass problem in Sec.~\ref{sec:he4}, and finally we conclude in Sec.~\ref{sec:conclusion}.

\section{Electron-nucleus scattering}
\label{sec:lepton-nuc_scattering}

Let us consider the process 
\begin{equation}
e(k) + A(p_0) \to e'(k') + f(p_f) \,,
\end{equation}
where an incoming electron with four-momentum $k=(E_k, \mathbf{k})$ is scattered off a nucleus $A$, producing an outgoing electron with four-momentum $k'=(E_k', \mathbf{k}')$ and a final (in general multi-particle) state $f$. 
The four-momentum transfer is $q\equiv(\omega, \mathbf{q})=k-k'$.
In the Born approximation, the electron interacts via the exchange of a single $\gamma$.

The inclusive cross section of this process can be written in terms of leptonic and hadronic tensors as
\begin{equation}
      \frac{d\sigma}{d \omega d\cos\theta} =\left(\frac{\alpha}{q^2}\right)^2 \frac{|\mathbf{k}|}{|\mathbf{k}'|} L_{\mu\nu} W^{\mu\nu} \,,
\end{equation}
with the angle of the outgoing lepton being $\theta$, $\alpha \approx 1/137$, is the fine structure constant. The lepton tensor is 
\begin{equation}
    L_{\mu\nu} = 2 [k_\mu k_\nu' + k_\mu' k_\nu -g_{\mu\nu} (kk') ] \,.
\end{equation}
The nuclear structure information is encoded in the hadron tensor
\begin{equation}
    \label{eq:hadronTens}
    W^{\mu\nu} = \sum_f \delta^4(p_0+q-p_f) \langle 0 | \left(J^\mu\right)^\dagger | \Psi_f \rangle \langle \Psi_f | J^\nu |0\rangle \,,
\end{equation}
where the current $J^\mu$ corresponds to the electromagnetic process.

In the following, we will focus on the quasi-elastic mechanism, for which the interaction takes place on a single nucleon, kicking off a nucleon from the remaining $(A-1)$ nucleus in the final state.
The electromagnetic current is a sum of one-body contributions which in the second quantization form is given by
\begin{equation}
   J^\mu =   \sum_{\alpha,\beta} \langle \beta| j^\mu |\alpha\rangle  a_{\beta}^\dagger a_{\alpha}\,.
   \label{eq:1bcurrent}
\end{equation}
Here, the initial and final nucleon states are labelled by $\alpha$ and $\beta$, respectively.
Within the spectral function formalism, we will use the fully relativistic current in the matrix element treating the initial and final nucleons as free states
\begin{equation}
    \langle p+q | j^\mu | p\rangle = \bar{u}( p+q)  V^\mu u( p) \ .
\end{equation}
The current $j^\mu$ has a vector structure and $u$ denotes a
Dirac spinor. 
Constructing the most general form of $V^\mu$ using the available four-vectors, we have
\begin{equation}
        V^\mu = F_1 \gamma^\mu + \frac{F_2}{2m}i\sigma^{\mu\nu} q_\nu\, .
\end{equation}
We use $F_1^{n,p}$, $F_2^{n,p}$ parametrized as in Ref.~\cite{Bradford:2006yz}.

\subsection{The Impulse Approximation}
\begin{figure}[hbt]
    \includegraphics[width=0.3\textwidth]{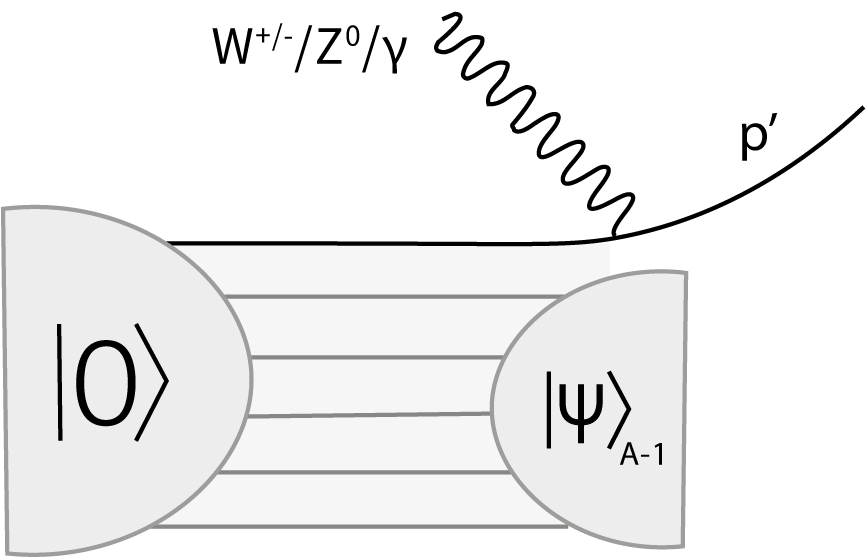}
  \caption{Impulse approximation: the electroweak interaction takes place on a single nucleon which afterwards does not interact with the spectator $|\Psi_{A-1}\rangle$ system.}
  \label{fig:ia}
\end{figure}

At relatively large momentum transfer $\mathbf{q}$, one can assume that the struck nucleon is decoupled from the nuclear $(A-1)$ system, i.e., that the final-state interaction can be neglected. Within this impulse approximation, the final nuclear state factorizes as
\begin{equation}
|\Psi_f\rangle \longrightarrow  a_{p'}^\dagger |\Psi_{A-1}\rangle \, ,
\label{eq:fact}
\end{equation}
where a plane-wave state $a_{p'}$ with momentum $\mathbf{p'}$ and energy $E_{p'}$ is added on top of the final $(A-1)$ system~\footnote{For simplicity we suppress spin and isospin indices.}. 
Using the current of Eq.~\eqref{eq:1bcurrent}, the one-body matrix element can be factorized as
\begin{align}
\label{ansatz}
\langle \Psi_f | J^\mu |0 \rangle &\to   \sum_{\alpha,\beta} \langle \beta| j^\mu |\alpha\rangle \langle\Psi_{A-1}| a_{p'} a_{\beta}^\dagger a_{\alpha}  |0\rangle \nonumber \\
&=  \sum_{\alpha,\beta} \langle \beta| j^\mu |\alpha\rangle \langle\Psi_{A-1}| \delta_{\beta\,p'} a_{\alpha}- a_{\beta}^\dagger a_{p'}  a_{\alpha}  |0\rangle \nonumber \\
&\approx \sum_\alpha \langle p'| j^\mu |\alpha \rangle  \langle\Psi_{A-1}|a_\alpha  |0\rangle \nonumber \\
&=\int \frac{d^3\mathbf{p}}{(2\pi)^3} \langle p'| j^\mu |p \rangle \sum_\alpha \langle \mathbf{p}|\alpha\rangle  \langle\Psi_{A-1}|a_\alpha  |0\rangle \,,
\end{align}
where the approximation in the third line assumes that the struck nucleon at the interaction vertex is exactly the one which is  ejected from the nucleus~\cite{mbte} and in the last line we insert a complete set of states $\int d^3p/(2\pi)^3 |p\rangle\langle p|$. $\langle \mathbf{p}|\alpha\rangle$ are single-particle wavefunctions in momentum space. The process is shown schematically in Fig.~\ref{fig:ia}.

The recoil energy $E_f^{kin}$ of the $(A-1)$ system is negligible for heavy nuclei, and the excitation is given by the energy conservation
\begin{equation}
E_{f}^{A-1}=\omega+E_0-E_{p'}-E_f^{kin}
\end{equation}
with the initial-state energy $E_0$. 

Substituting Eq.~(\ref{ansatz})  in Eq.~\eqref{eq:hadronTens}, the incoherent contribution to the hadron tensor becomes

\begin{align}
W^{\mu\nu}(q)=& \int \frac{d^3\mathbf{p} d^3\mathbf{\tilde p}}{(2\pi)^6} \sum_{\alpha,\alpha'}\sumint_{\Psi_{A-1}} \, \langle p | \left(j^{\mu}\right)^\dagger |p^\prime \rangle \langle p^\prime |  j^{\nu} | \tilde{p} \rangle\nonumber\\
&\times \langle \mathbf{p}|\alpha\rangle^\dagger \langle \mathbf{\tilde p}|\alpha'\rangle \langle 0|a_{\alpha}^\dagger  |\Psi_{A-1}\rangle  \langle\Psi_{A-1}|a_{\alpha'}  |0\rangle \nonumber  \\
&\delta(\omega-E_{p'} - E_{f}^{A-1} - E_{f}^{kin} +E_0)\,\, .
\end{align}
From the momentum conservation at the single nucleon vertex $\mathbf{p}=\mathbf{\tilde{p}}=\mathbf{p^\prime}-\mathbf{q}$. Furthermore, the spin state of $a_\alpha$ and $a_{\alpha'}$ coincide due to charge conservation 
and the assumption that the nuclear ground state has spin zero. Finally, the last step of the factorization separates the energy conservation at the vertex from the excitation of the residual system by introducing the energy $E$ needed to remove a nucleon with momentum $\mathbf{p}$ from the ground state as
\begin{align}
&\delta(\omega+E_0-E_{p'} - E_{f}^{A-1}-E_f^{kin} ) \nonumber \\
&= \int dE \delta (\omega+E-E_{p'}-E_f^{kin}) \delta(E+E_{f}^{A-1}-E_0)\, .
\end{align}
Using this equation and introducing explicitly the isospin dependence,  the hadron tensor is
\begin{align}
\label{eq:hadron_tens_SF}
W^{\mu\nu}(q)&= \int \frac{d^3\mathbf{p}}{(2\pi)^3} dE \big[ S^n(\mathbf{p},E)w_{n}^{\mu\nu}(p,q)+ \nonumber \\
& S^p(\mathbf{p},E)w_{p}^{\mu\nu}(p,q) \big] \delta(\omega+E-E_{p+q}-E_f^{kin})\, ,
\end{align}
with $w_{n,p}^{\mu\nu}(p,q) = \langle p+q| j^{\mu} |p\rangle^\dagger \langle p+q | j^{\nu} |p\rangle$ depending on the isospin of $|p \rangle$, where we have introduced the hole spectral function
\begin{align}
S^{n,p}(\mathbf{p},E)&=\sum_{\alpha,\alpha'}\sumint_{\Psi_{A-1}}  |\langle 0|a_\alpha^\dagger  |\Psi_{A-1}\rangle \langle \Psi_{A-1}| a_{\alpha'}| 0\rangle  \nonumber \\
&\langle \mathbf{p}|\alpha\rangle^\dagger \langle \mathbf{p}|\alpha'\rangle \delta(E+E_{f}^{A-1}-E_0)\,.
\label{eq:SF_def}
\end{align}
The  spectral function gives the probability distribution of removing a nucleon with momentum ${\mathbf p}$ from the target nucleus, 
leaving the residual $(A-1)$ system with an energy $E_0-E$. For closed-shell nuclei, such as the $^4$He considered in this work, the  spectral functions of spin-up and spin-down nucleons coincide.  We normalize spectral functions as
\begin{equation}
    \int \frac{d^3p}{(2\pi)^3} dE S^{n(p)}(\mathbf{p},E) = N(Z)\,.
\end{equation}
In the relativistic regimes, the factors  $m/E_{p}$ and $m/E_{p+q}$ should be included to account for the implicit covariant normalization of the four-spinors of the nucleons in the matrix elements of the current $j^\mu$. Hence the hadron tensor finally becomes
\begin{align}
W^{\mu\nu}(q)&= \int \frac{d^3\mathbf{p}}{(2\pi)^3} dE \frac{m}{E_p}\frac{m}{E_{p+q}}\nonumber \\
&\big[ S^n(\mathbf{p},E)w_{n}^{\mu\nu}(p,q)+  S^p(\mathbf{p},E)w_{p}^{\mu\nu}(p,q) \big]\nonumber \\
&\times \delta(\omega+E-E_{p+q}-E_f^{kin})\, ,
\label{eq:wmunu-IA}
\end{align}
where one can see that it can be calculated starting from the spectral function for $(n)$ neutrons and $(p)$ protons.

We performed the factorization of the relativistic currents and the nuclear ground state governed by non-relativistic dynamics. This way we can address the processes occurring at high energy-momentum transfers. This procedure introduces, however, some model dependence since we do not treat the wavefunction and the currents on an equal footing.
In a consistent description we can either have the picture of a simple current interacting with a complicated nucleus, or an alternative picture (and a continuum of approaches between these two extremes) where the nucleus is simple (e.g., a product state), and the current is complicated and consists of one- and two-body terms. Very recently, authors of Ref.~\cite{tropiano2021} presented a detailed discussion of this subject.  
In particular they analyse how the high-momentum behaviour of the wavefunctions depends on the resolution of employed nuclear Hamiltonian. Their results can be applied to the momentum distribution of the spectral functions. 
We leave the analysis of this effect, as well the role of two-body currents within the factorization scheme, for the future studies.

%
%
%
%
%
%
%
\section{Formalism}
\label{sec:formalism}
\subsection{Green's function and spectral function}
The spectral functions Eq.~\eqref{eq:SF_def} are defined through the imaginary part of a propagator in a many-body system. Presently, we will consider only the hole propagation of a state with quantum numbers $\alpha$ to a state with quantum numbers $\beta$
\begin{equation}
\label{eq:Green}
\begin{split}
    &G_h(\alpha,\beta, E)  = \langle 0| a_\beta^\dagger \frac{1}{E-(E_0-\hat H)-i\epsilon}  a_\alpha |0\rangle\,,\\
    &\mathrm{Im}G_h(\alpha,\beta, E) =\\
    &-\pi \sumint_{\Psi_{A-1}} \langle 0| a_\beta^\dagger |\Psi_{A-1} \rangle \langle \Psi_{A-1}|  a_\alpha |0\rangle \delta\big(E-(E_0-E_{\Psi})\big)\,.
\end{split}
\end{equation}
The spectral functions can be retrieved from the imaginary part of Green's function summing over all the appropriate quantum numbers
\begin{equation}
S(\mathbf{p},E) = -\frac{1}{\pi} \sum_{\alpha, \beta} \langle \mathbf{p}|\alpha\rangle \langle \mathbf{p}|\beta\rangle^\dagger \mathrm{Im} G_h(\alpha,\beta,E)\,,
\label{eq:SF}
\end{equation}
The reconstruction of $\mathrm{Im}G_h(\alpha,\beta, E)$ from Eq.~\eqref{eq:Green} requires a summation over all the excited states $|\Psi_{A-1}\rangle$, which contains not only  bound states but also continuum states. Within ab-initio methods the calculated spectrum is typically discretized because of the truncation of the many-body space. 
Continuum effects can be included via complex scaling techniques~\cite{Suzuki:2005wv, Carbonell:2013ywa} or in the Berggren basis. The latter idea has been recently applied to obtain the microscopical optical potential from the coupled-cluster theory~\cite{Rotureau:2016jpf,Rotureau:2018pxk}.
These techniques are usually combined with the Lanczos algorithm to construct tri-diagonal forms of large matrices, and thereby give access to the extreme eigenvalues of the problem.
Here, however, we will use another approach described in the next section.

\subsection{Chebyshev expansion of integral transform}
\label{sec:ChEK}
Within the ChEK method we rephrase our problem: instead of reconstructing the response we want to estimate observables which are expressed as the energy integrals of the response.
The method can be used in a general situation
\begin{equation}
\label{eq:phi}
\Phi = \int d\omega f(\omega) R(\omega)\,,
\end{equation}
where $f(\omega)$ is any bound function defining the observable and $R(\omega)$ is a response function -- which in our case  corresponds to $\mathrm{Im}G_h(\alpha,\beta,\omega)$. 
Our strategy to approximate the quantity in Eq.~\eqref{eq:phi} consists in applying the integral transform $\tilde R(E)$
\begin{equation}
\tilde\Phi = \int d E f(E) \tilde R(E)\,,
\end{equation}
in such a way that we control the approximation error $|\Phi-\tilde\Phi|$. Let us also notice, that the reconstruction of $\Phi$ does not require the inversion of integral transform.
In our case $\tilde R(E)$ is given by
\begin{equation}
\begin{split}
    \mathrm{Im}\tilde G_h(\alpha,\beta, E) &\\
    &=\int d\omega \mathrm{Im}G_h(\alpha,\beta, \omega) K(\omega,E)\\
    &=-\pi \sumint_{\Psi_{A-1}}  \langle 0| a_\beta^\dagger |K\big(E_{\Psi},E-E_0\big) |  a_\alpha |0\rangle \\
    &=-\pi\langle 0| a_\beta^\dagger |K\big(\hat H,E-E_0\big) |  a_\alpha |0\rangle \,.
\end{split}
\label{eq:imGreen_IT}
\end{equation}
The kernel $K(\omega,E)$  can be realized by various functions. 
Here, we will apply the Gaussian kernel
\begin{equation}
K(\omega, E; \lambda) = \frac{1}{\sqrt{2\pi} \lambda} \exp\left(-\frac{(\omega-E)^2}{2\lambda^2}\right)\,.
\end{equation}
Following  Ref.~\cite{Roggero:2020qoz}, we  characterize the kernel as $\Sigma$-accurate with $\Lambda$-resolution:
\begin{equation}
\label{eq:kernel_char}
    \sup_{\omega_0\in[-1,1]}\sumint_{\omega_0-\Lambda}^{\omega_0+\Lambda} K(\omega_0,E) dE \geq 1-\Sigma\, .
\end{equation}
With these definitions we can provide the uncertainty bound for $|\Phi-\tilde\Phi|$, which will depend on the properties of the function $f$ and the kernel $K$.

Next we expand Eq.~\eqref{eq:imGreen_IT} into Chebyshev polynomials and truncate the number of terms $N$. This truncation will introduce an additional error, as will be explained later. The truncated kernel
\begin{equation}
    K(\omega,E) = \sum_{k=0}^N c_k(E) \mathcal{T}_k(\omega)\,,
    \label{eq:kernel_expansion}
\end{equation}
is expressed in terms of $\mathcal{T}_k(\omega) = \cos [k\arccos(\omega) ]$ which follow a recursive relation
\begin{equation}
\begin{split}
 &\mathcal{T}_0(x) = 1;\, \, \, \, \,\mathcal{T}_{-1}(x) = \mathcal{T}_1(x) = x;\\
 &\mathcal{T}_{k+1}(x) = 2x \mathcal{T}_k(x) - \mathcal{T}_{k-1}(x)\,.
\label{eq:chebyshev}   
\end{split}    
\end{equation}

Let us assume that the Hamiltonian norm is known, and that we are able to rescale our problem $[E_{min}, E_{max}] \rightarrow [-1,1]$. This allows us to use Chebyshev polynomials (which are defined on the interval $[-1,1]$). The Hamiltonian spectrum can be obtained, e.g., via the Lanczos algorithm. The rescaling is given then by:
\begin{equation}
\label{eq:rescaling}
    \begin{split}
        &a = (E_{max} - E_{min})/2\, ,\ \ \ \ \ b = (E_{max} + E_{min})/2 \\
        & H:= (H-b)/a\, .
    \end{split}
\end{equation}
Combining Eqs.~\eqref{eq:imGreen_IT} and~\eqref{eq:kernel_expansion} we obtain
\begin{equation}
\begin{split}
    \mathrm{Im}\tilde G_h(\alpha,\beta, E) &\\
    =&-\pi \sum_{k=0}^N c_k(E)  \langle 0| a_\beta^\dagger \mathcal{T}_k\big(\hat H\big)   a_\alpha |0\rangle \\
    \equiv&-\pi \sum_{k=0}^N c_k(E) \mu_k \,.
\end{split}
\label{eq:imGreen_kernel}
\end{equation}
For simplicity we will abuse the notation and understand that $\mathrm{Im}\tilde G_h(\alpha,\beta, E)$ has an implicit $N$ dependence. Furthermore, the moments $\mu_k$ have an implicit dependence on $\alpha$ and $\beta$.
The moments of the expansion $\mu_k$ can be retrieved from a many-body calculation, using the recursive relation from Eq.~\eqref{eq:chebyshev}
\begin{equation}
\label{eq:cheb_rec}
\begin{split}
    & \langle \tilde \Psi_0| \equiv \langle 0| a_\beta^\dagger| \, ,\ \ \ \ \ |\Psi_0 \rangle \equiv a_\alpha |0\rangle\, ,\\
    &\langle \tilde \Psi_k| \equiv\langle \tilde \Psi_{k-1}| \hat H \ \ \ \ \ |\Psi_k\rangle = \hat H |\Psi_{k-1}\rangle\\
    & \mu_0 =  \langle \tilde \Psi_0|\Psi_0 \rangle \, ,\ \ \ \ \ \mu_1 =   \langle \tilde \Psi_0| \Psi_1 \rangle \equiv \langle \tilde \Psi_1| \Psi_0 \rangle \\
    & \mu_{k+1} = 2 \langle \tilde\Psi_0| \Psi_{k+1} \rangle -\mu_{k-1}\equiv 2 \langle \tilde \Psi_{k+1}| \Psi_0 \rangle-\mu_{k-1}\,.
\end{split}    
\end{equation}
In the $(k+1)$-th step  only the $|\Psi_{k}\rangle$ (or $\langle \tilde \Psi_{k}|$)  state has to be known from the previous iteration. Similarly to the Lanczos procedure, we iterate the action of Hamiltonian $\hat H$. Here, however, no orthogonality restoration is needed at each step, which makes the procedure faster and requires less memory.
The coefficients $c_k$ from Eq.~\eqref{eq:imGreen_kernel} depend on the chosen kernel and their form can be found in Ref.~\cite{Sobczyk:2021ejs}.

In the present case, we will define the function $f(\omega)$ in Eq.~\eqref{eq:phi} as a histogram bin centered at $\eta$ and a half-width $\Delta$
\begin{equation}
\label{eq:window}
 f(\omega) \equiv h_{\Delta} (\eta, \omega) = \bigg\{\begin{matrix}
0 & |\eta-\omega|>\Delta\\
1 & \text{otherwise}
\end{matrix}\;.
\end{equation}
We are then interested in approximating the histogram
\begin{equation}
\begin{split}
    \mathrm{Im}G_h(\alpha,\beta; \eta, \Delta) = \int h_{\Delta}(\eta,E) \mathrm{Im}G_h(\alpha,\beta, E) dE
\end{split}
\label{eq:imGreen_hist}
\end{equation}
using its integral transform (given in Eq.~\eqref{eq:imGreen_kernel}) with a finite number of Chebyshev moments $N$
\begin{equation}
\begin{split}
    \mathrm{Im}\tilde G_h(\alpha,\beta; \eta, \Delta) =  -\pi \sum_{k=0}^N \mu_k \int h_{\Delta}(\eta,E) c_k(E) dE\,.
\end{split}
\label{eq:imGreenIT_hist}
\end{equation}
As  shown in Ref.~\cite{Sobczyk:2021ejs}, we get
\begin{equation}
\label{eq:error_est}
    \begin{split}
 \mathrm{Im}\tilde G_h(\Delta -&\Lambda)-\Sigma-2\gamma(\Delta -\Lambda)  \\
 &\leq\mathrm{Im}G_h(\Delta )  \\
 \leq \mathrm{Im}\tilde G_h&(\Delta +\Lambda)+\Sigma+2\gamma(\Delta +\Lambda)\,,
    \end{split}
\end{equation}
where we used a shorter notation $\mathrm{Im}G_h(\alpha,\beta; \eta, \Delta)\equiv \mathrm{Im}G_h(\Delta)$. The truncation error $\gamma$ depends on the number of moments $N$ and the properties of the kernel
\begin{equation}
    \gamma = \sum_{k=N}^\infty c_k(E) \mathcal{T}_k(\omega)\,.
    \label{eq:beta}
\end{equation}
The analytical expression for the bounds on $\gamma$ can be found in Eqs.~(B5),~(B22) of Ref.~\cite{Sobczyk:2021ejs} for the Lorentzian and Gaussian kernels, respectively. As has been advocated in Ref.~\cite{Sobczyk:2021ejs}, the Gaussian integral transform has better convergence properties and will  therefore be used in the present calculations.

Eq.~\eqref{eq:error_est} is the master equation which will ultimately allow us to reconstruct the spectral functions defined in Eq.~\eqref{eq:SF} as a histogram. It gives the error bound for $\mathrm{Im}G_h(\Delta)$ depending on the characteristics of the kernel, $\Lambda$ and $\Sigma$, and the number of Chebyshev moments $N$ (which enter both $\gamma$ and $\mathrm{Im}\tilde G_h(\Delta\pm\Lambda)$). 

It is important to notice some properties of the integral transform $\mathrm{Im}\tilde G_h(\Delta\pm\Lambda)$ (see Eq.~\eqref{eq:imGreenIT_hist}). The characteristics of the kernel is encoded in coefficients $c_k$. In this way, the Chebyshev moments $\mu_k$ have to be calculated only once for any kernel to be used. This is an important feature, because their computation is much more expensive than the post-processing stage (i.e. constructing histograms). Moreover, the integral of Eq.~\eqref{eq:imGreenIT_hist} can be performed  analytically for the Gaussian kernel, which speeds up the calculation and does not introduce any additional numerical errors.

%
%
%
%
%
%
%

\subsection{Coupled-cluster theory}

The moments of the Chebyshev expansion $\mu_k$ in Eq.~\eqref{eq:cheb_rec} have to be calculated in a many-body framework. In this work we employ the spherical coupled-cluster method~\cite{hagen2014}, which can accurately describe ground- and excited state properties of nuclei in the neighbourhood of closed (sub-)shell nuclei. The method starts from a spherical Hartree-Fock reference state $|\Psi_\mathrm{HF}\rangle$ and includes correlations with an exponential ansatz
\begin{equation}
    |0\rangle = e^T |\Psi_\mathrm{HF}\rangle \,.
\end{equation}
Here the cluster operator $T$ is built of 1p-1h, 2p-2h, $\dots$ excitations
\begin{equation}
\label{eq:excitationOperator}
    T=\sum_{i,a} t^a_i a^\dagger_a a_i + \frac{1}{4}\sum_{ijab}t^{ab}_{ij}a_a^\dagger a_b^\dagger a_i a_j +...\,,
\end{equation}
and is truncated at a certain level. In this work we truncate $T$ at the 2p-2h excitation level, which is known as the coupled-cluster singles-and-doubles (CCSD) approximation.\footnote{In Eq.~\eqref{eq:excitationOperator} index $a$, $b$ iterates over particle states, while $i$, $j$ over hole states.} The amplitudes $t$ are obtained by solving a large set of coupled non-linear equations, which are subsequently used in the construction of the similarity transformed Hamiltonian and creation/annihilation operators
\begin{equation}
    \overline{H} = e^{-T} \hat{H} e^T,\ \ \ \ \overline{a_\alpha} = e^{-T} a_\alpha e^T, \ \ \ \  \overline{a^\dagger_\alpha} = e^{-T} a^\dagger_\alpha e^T\\.
\end{equation}

For our problem, the initial states are built as
\begin{equation}
\begin{split}
&    |\Psi_0\rangle = \overline{a_\alpha} |\Psi_\mathrm{HF}\rangle\, , \ \ \ \ \ \langle\tilde\Psi_0| = \langle \Psi_\mathrm{HF} | \overline{a_\beta^\dagger}\, .
\end{split}
\end{equation}
The calculation of Chebyshev moments follows Eq.~\eqref{eq:cheb_rec}, which requires iterating
\begin{equation}
\begin{split}
&    |\Psi_k\rangle = \overline{H} |\Psi_{k-1}\rangle\, , \ \ \ \ \ \langle\tilde\Psi_k| =  \langle\tilde\Psi_{k-1}|\, \overline{H} 
\end{split}
\end{equation}
The action of the Hamiltonian can be accumulated either on the right state, or on the left, or distributed between them. This allows for a numerical check of the procedure.

In our calculation we use the NNLO$_\mathrm{sat}$ nucleon-nucleon and three-nucleon interaction, which was adjusted to the binding energy and charge radii of light nuclei and selected oxygen isotopes~\cite{Ekstrom:2015rta}. Furthermore, we approximate the three-nucleon interaction at the normal-ordered two-body level which has been shown to be accurate for light- and medium mass nuclei~\cite{hagen2007a,roth2012}. We note that this approximation breaks translational invariance of the Hamiltonian, and impacts the computation of intrinsic observables in light nuclei~\cite{djarv2021}.
The results for various observables are converged in a model space of 15 oscillator shells ($N_\mathrm{max}=14$) using the oscillator spacing $\hbar\omega=16$~MeV. The three-nucleon interaction had an additional energy cut on allowed configurations given by $E_{3\rm max} = 16$~MeV. 
%

%
%
%
%
%
%

\section{Results}
\label{sec:he4}

Before the analysis of the spectral function itself, we benchmark our result for the momentum distribution, which is directly derived from the spectral function:
\begin{equation}
    n(\mathbf{p}) = \int dE\,  S(\mathbf{p},E) = \sum_{\alpha,\beta} \langle \mathbf{p}|\alpha\rangle \langle \mathbf{p}|\beta\rangle^\dagger \langle 0|a^\dagger_\beta a_\alpha|0\rangle \,.
\end{equation}

\begin{figure}[hbt]
    \includegraphics[width=0.5\textwidth]{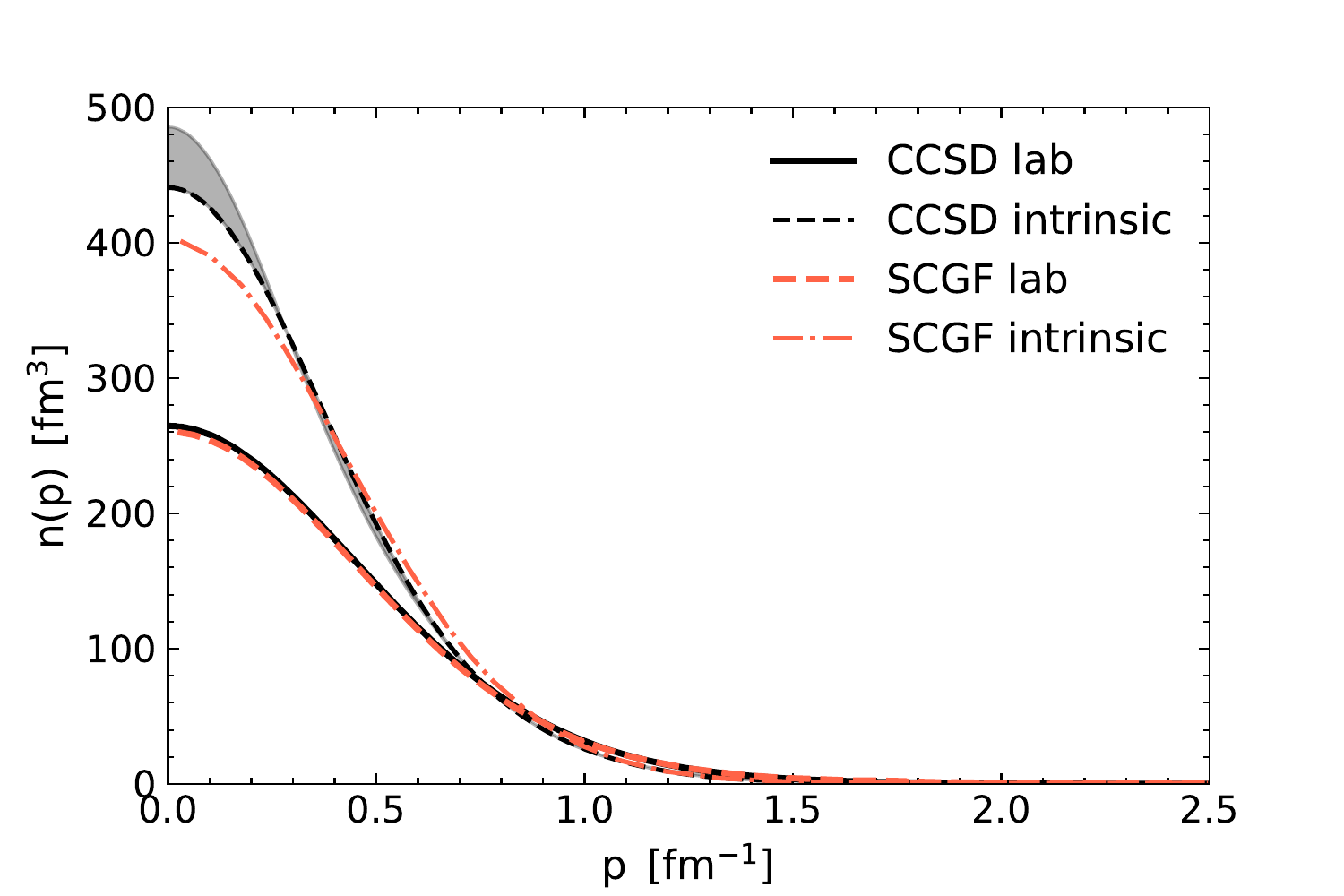}
  \caption{Comparison of laboratory and intrinsic momentum distribution in $^4$He (solid and dashed black lines). The grey band corresponds to an uncertainty of our procedure of removing the center of mass wavefunction. Results of SCGF~\cite{Rocco:2018vbf} are shown for the laboratory (dashed line) and intrinsic momentum (dashed-dotted line). All the calculations were done with NNLO$_\mathrm{sat}$ interaction and the same model space.}
  \label{fig:mom_comparison}
\end{figure}

As coupled-cluster computations are performed in the laboratory system, one has to extract the intrinsic spatial density (or intrinsic momentum distribution) from the corresponding laboratory distributions.
Within the coupled-cluster and in-medium similarity renormalization group frameworks the nuclear ground-state has been shown to factorize with a very good precision into an intrinsic wavefunction and a Gaussian center-of-mass~\cite{Hagen:2009pq,hagen2010b,parzuchowski2017} when the kinetic energy of the center of mass is removed from the Hamiltonian. We note that this factorization was demonstrated in the coupled-cluster approach for a two-body Hamiltonian, while the in-medium similarity renormalization group approach showed that this factorization holds when applying a three-nucleon interaction in the normal-ordered two-body approximation in nuclei as light as $^{14}$C \cite{parzuchowski2017}. 
Since the effect of breaking translational invariance was found to be small on the binding energy and radius of $^4$He~\cite{Ekstrom:2015rta}, we assume that factorization holds also for $^4$He in the coupled-cluster theory. 
Assuming that the center of mass is a Gaussians, the extraction of intrinsic momentum distribution involves a deconvolution via Fourier transforms, and  details are presented in the Appendix~\ref{sec:appendix}. Because of numerical reasons---varying the cut-off in the Fourier transform---the low-momentum region is affected by a few-percent uncertainty.

Figure~\ref{fig:mom_comparison} shows the intrinsic and laboratory proton momentum distributions computed within CCSD.  The difference is clearly visible for low and intermediate momenta up to $k\approx 0.7\mathrm{fm}^{-1}$. 
We compare our results with those from the SCGF method. While the results coincide for the laboratory momentum distribution, there are visible differences to the intrinsic CCSD momentum density. We mostly ascribe them to two very different strategies of the center-of-mass removal. In our case the method is straightforward and relatively simple, while the procedure employed in Ref.~\cite{Rocco:2018vbf} consists of two steps. Firstly the SCGF result is approximated via an optimized reference state, then the center-of-mass component is removed from the wavefunction using Monte Carlo Metropolis sampling. We note that the three-nucleon interaction is approximated slightly different in the SCGF approach~\cite{carbone2013}, and may therefore impact the comparison with our approach for intrinsic observables in $^4$He. 
We speculate that this difference will lead to some discrepancies in the cross-section predictions. However, the low-momentum region plays a minor role since the hadron tensor is weighted by $p^2dp$ [see Eq.~\eqref{eq:wmunu-IA}].

\subsection{Spectral function}
Benchmarking of the momentum distribution $n(\mathbf{p})$ allows us to validate the momentum dependence of the spectral function $S(\mathbf{p},E)$ and compare it with a previous calculation of Ref.~\cite{Rocco:2018vbf}. 
However, the  spectral function energy dependence requires a more careful analysis. The energy distribution, driven by $\mathrm{Im}G(\alpha,\beta,E)$ is obtained via the integral transform expanded into Chebyshev polynomials. These are calculated according to recursive relations of Eq.~\eqref{eq:cheb_rec} iterating the action of the Hamiltonian on the initial pivot state
\begin{equation}
    |\Psi_n\rangle = \hat H^n|\Psi_0\rangle = \hat H^n\, a_\alpha |0\rangle \ .
\end{equation}
Two remarks are in order. The initial state $|\Psi_0\rangle$ is composed of $(A-1)$ nucleons and, to be consistent, the Hamiltonian applied in the iteration should be changed accordingly. Otherwise the energy conservation of Eq.~\eqref{eq:Green}, $E_0-E_\Psi$ would be shifted since $E_\Psi$ is the excitation of the $(A-1)$ system. Additionally, $|\Psi_0\rangle$ might contain spurious center-of-mass excitations which should be detected and removed.
The first point was already discussed in Ref.~\cite{Hagen:2010zz} and a method to account for this inconsistency was proposed. It consists in performing the calculation of the ground state and excitation energies both for $A$ nucleons ($E_{0}$ and $\omega =E_0-E_\psi$ accordingly) and $(A-1)$ ($E_{0}^*$ and $\omega^*=E_0^*-E_\psi^*$ accordingly), so that we can calculate
\begin{equation}
    E_0-E_\Psi^* = E_0-E_0^*+\omega^* \ .
\end{equation}
We have checked that the difference between this value and $E_0-E_\Psi$ is around $1.5$ MeV. For the purposes of the  spectral function---which is a valid approximation for the momentum transfer of several hundreds of MeV and energy transfers of tens of MeV---this difference is not drastic. It will be also partially taken into account since we consider  spectral function in form of a histogram, whose binning will be larger than $1.5$ MeV.

\begin{figure}[hbt]
    \includegraphics[width=0.5\textwidth]{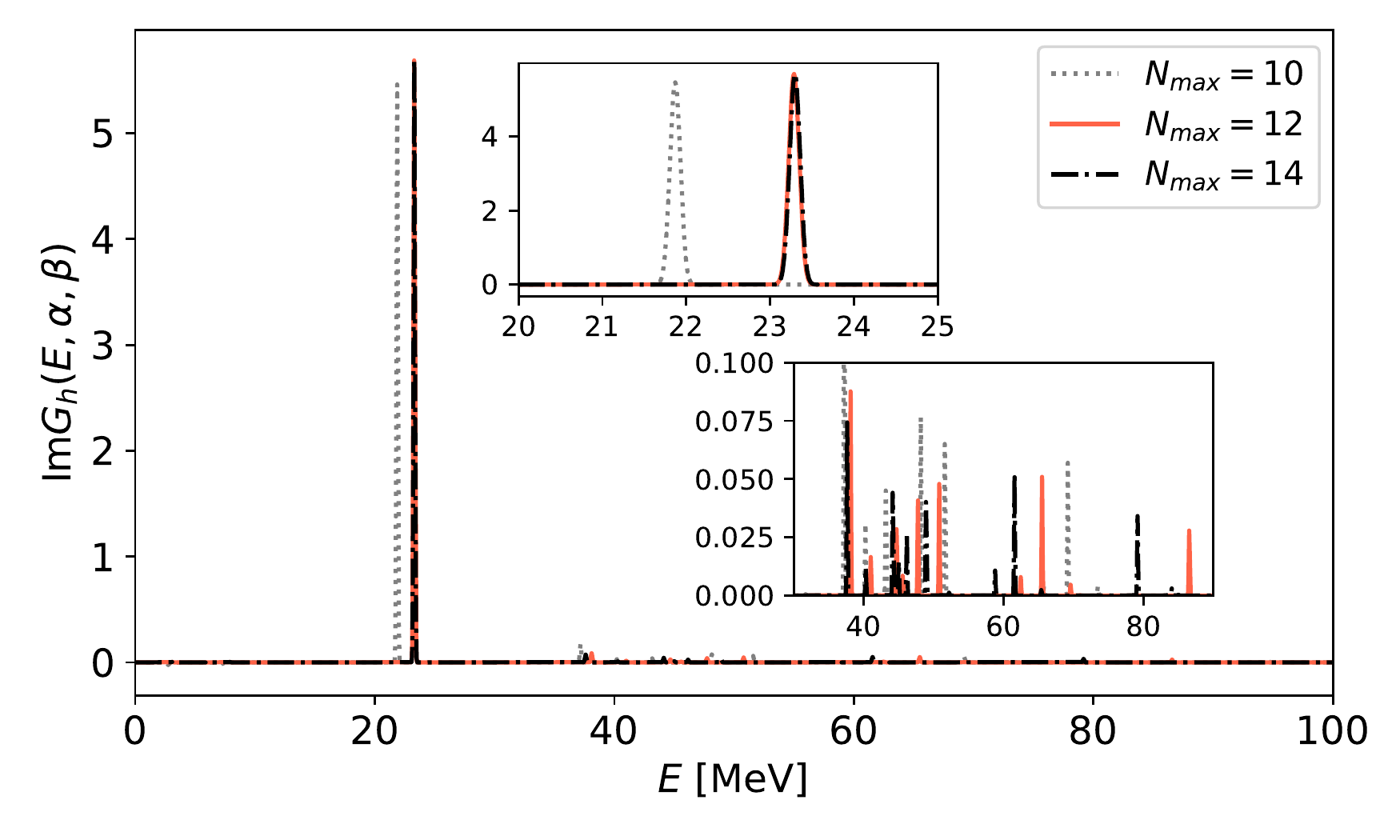}
  \caption{$\text{Im} G_h(E,\alpha,\beta)$ for $\alpha=\beta$ of $^4$He corresponding to a proton with the radial quantum number $n=0$, and orbital angular momentum $l=0$. }
  \label{fig:energy_convergence}
\end{figure}

The disentanglement of the center of mass from the physical excitations in $\mathrm{Im}G(\alpha,\beta,E)$ is complicated. However, the spectral function of $^4$He has a simple structure. It is dominated by a single peak whose position corresponds to the energy difference between the ground states of $^4$He and $^3$H (in the case of the proton  spectral function) or $^4$He and $^3$He (in the case of the neutron  spectral function). In Fig.~\ref{fig:energy_convergence}, we show how the convergence of $\mathrm{Im}G(\alpha,\beta,E)$ depends on $N_\mathrm{max}$  for the specific case of quantum numbers $\alpha=\beta$, radial quantum number $n=0$ and orbital angular momentum $l=0$. The dominant peak at around $23.5$~MeV is converged already for $N_{\mathrm{max}}=14$. The smaller excitations visible at higher energies play a negligible role. Their integrated strength does not exceed $2\%$. We also observe some very small contribution of states with negative strengths, which can be treated as unphysical excitations. We remove them from the final spectral function.

The analysis for $^4$He shows that our treatment, although introducing some approximations, still gives reasonable spectral functions within a few percent of uncertainty. We were able to remove the center-of-mass contamination from the momentum-dependent part of the  spectral function, and performed various checks to make sure that the center-of-mass excitation does not strongly affect the energy distribution. For heavier nuclei the situation is known to be better, since center-of-mass effects scale as $1/A$.

To obtain the final spectral function within the ChEK method we need to know the scaling factors, see Eq.~\eqref{eq:rescaling}, which we estimate through the Lanczos algorithm, $E_{min}=0$ and $E_{max}=200$~MeV. The reconstruction using histograms requires also to set the bin width, and $3$~MeV is a resolution that encompasses some of the earlier discussed uncertainties. Next, the Gaussian kernel width is chosen such that $\Lambda$ and $\Sigma$ are kept small (see Eq.~\eqref{eq:kernel_char}), while the number of Chebyshev moments required is still manageable. In the results presented in this work we used the Gaussian width $\lambda=0.12$~MeV and $\Lambda=0.5$~MeV. According to the results of Ref.~\cite{Sobczyk:2021ejs} the number of required moments would exceed $N=10000$ to keep the truncation error below $1\%$. While this number seems large it is still achievable within the CCSD approximation. Still, as was also pointed out in Ref.~\cite{Sobczyk:2021ejs}, the current bound on the truncation error is greatly overestimated. We have checked numerically that results are converged already with $N=6000$ moments.
With this choice of parameters we obtain a histogram spectral function whose errors are negligible. 

\subsection{Electron-nucleus scattering}
\begin{figure*}[hbt]
    \includegraphics[width=0.32\textwidth]{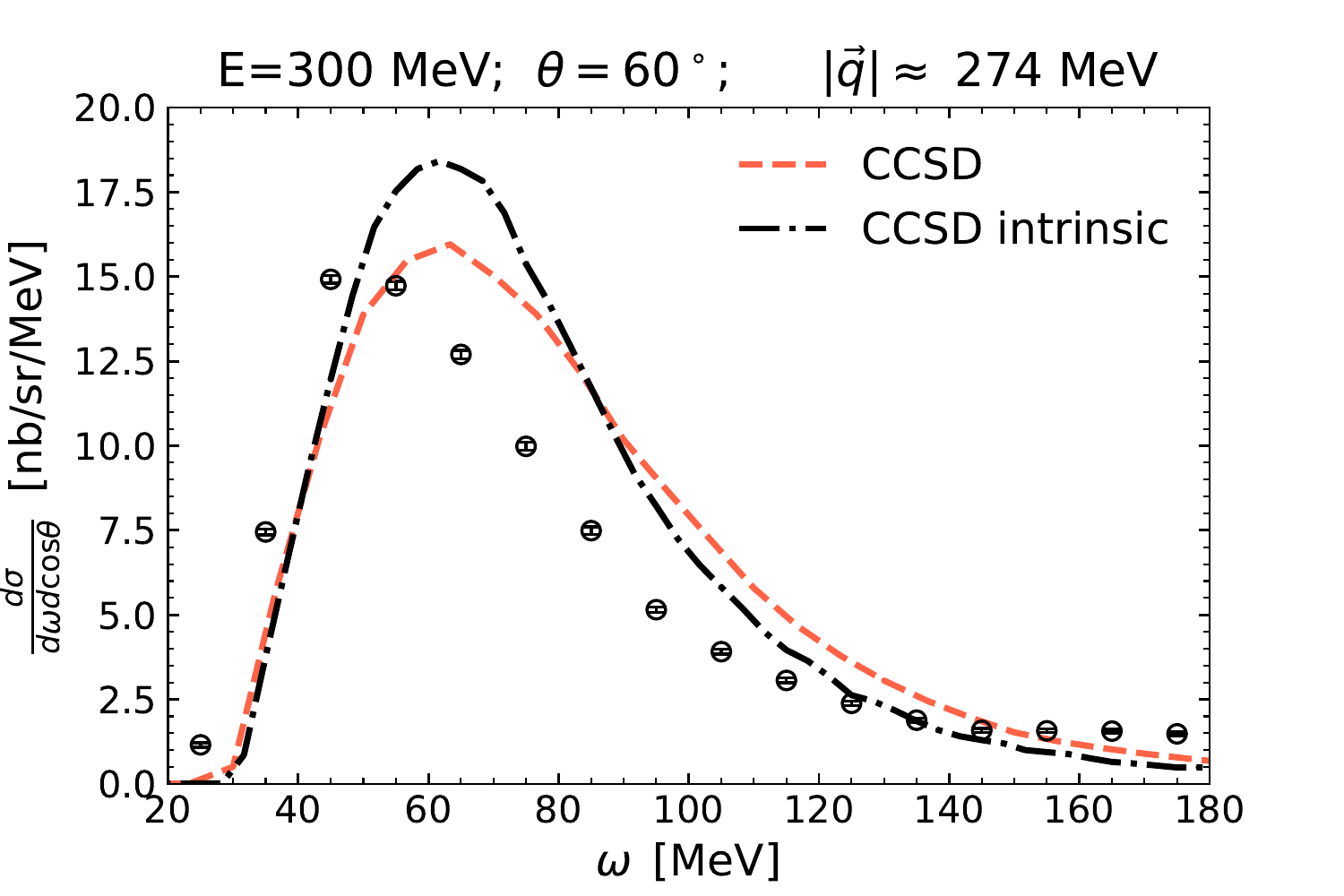}
    \includegraphics[width=0.32\textwidth]{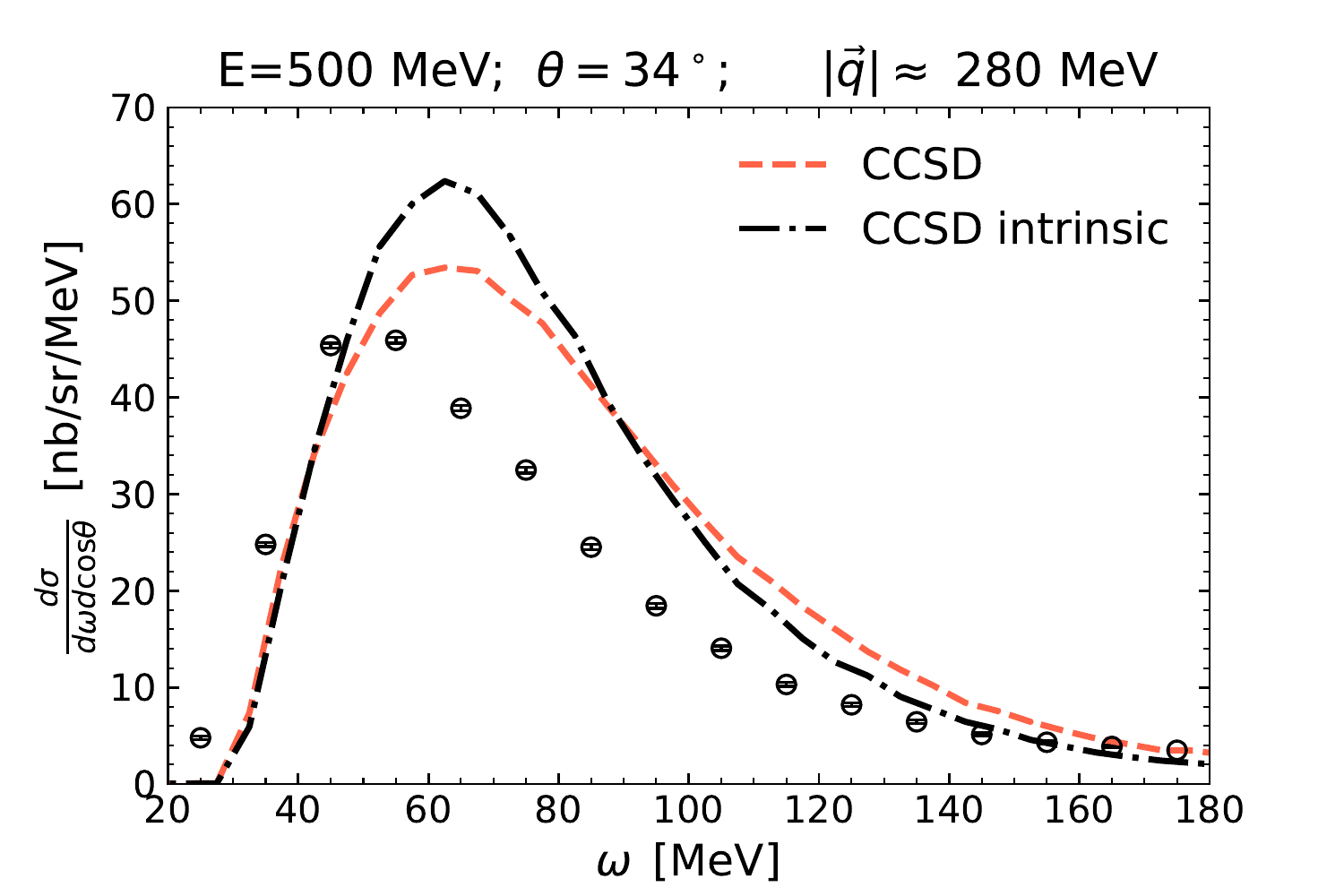}
    \includegraphics[width=0.32\textwidth]{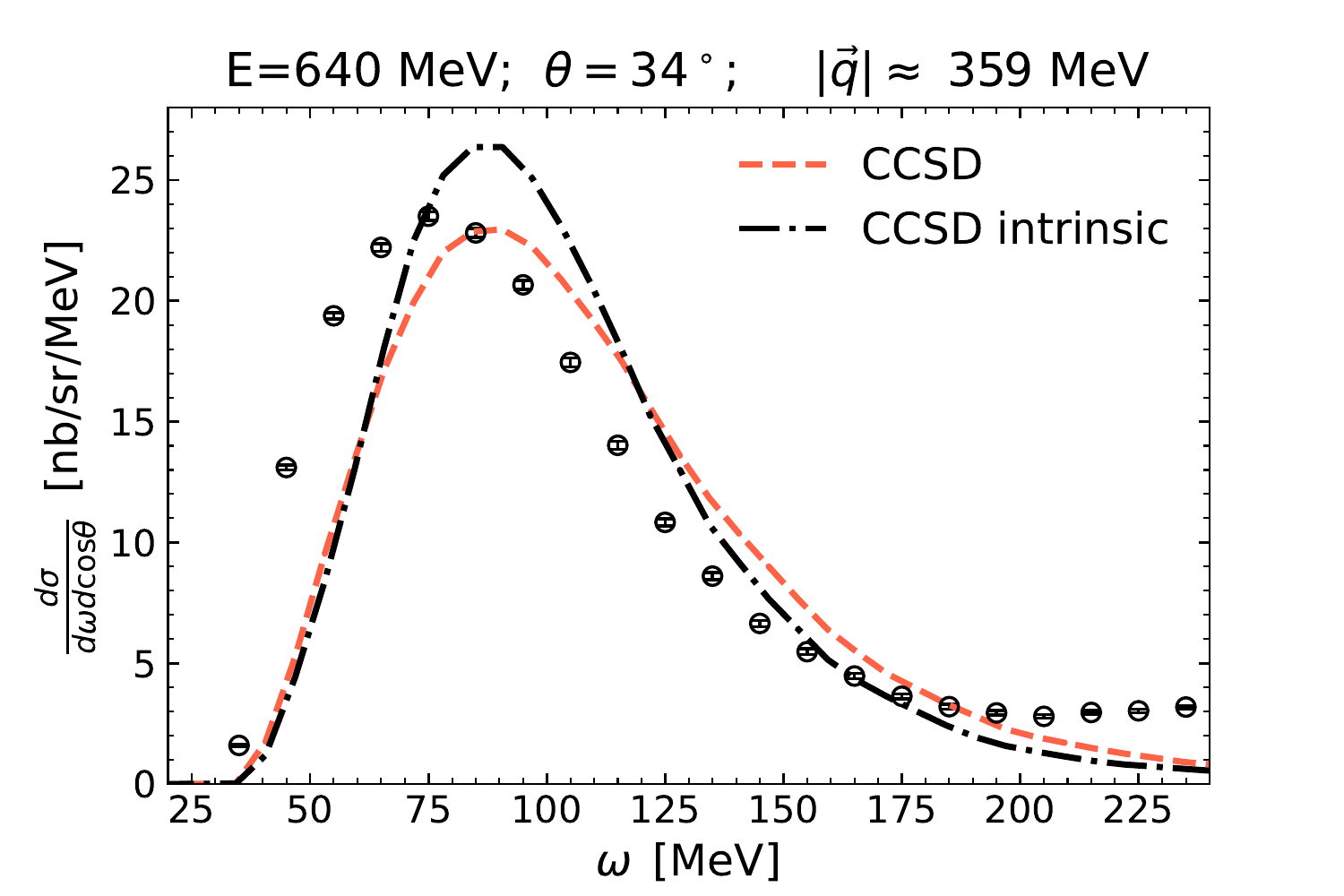}
    \includegraphics[width=0.32\textwidth]{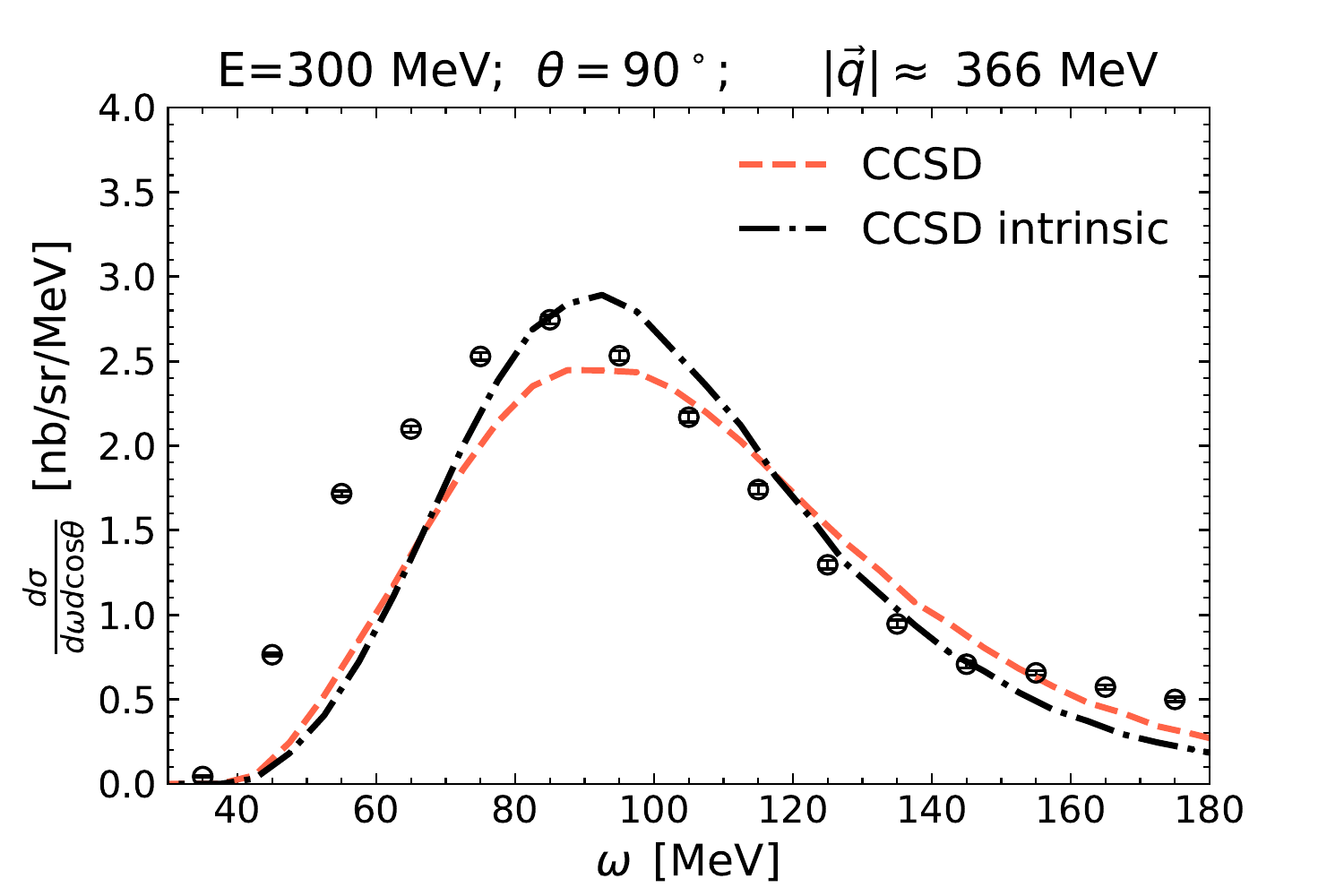}
    \includegraphics[width=0.32\textwidth]{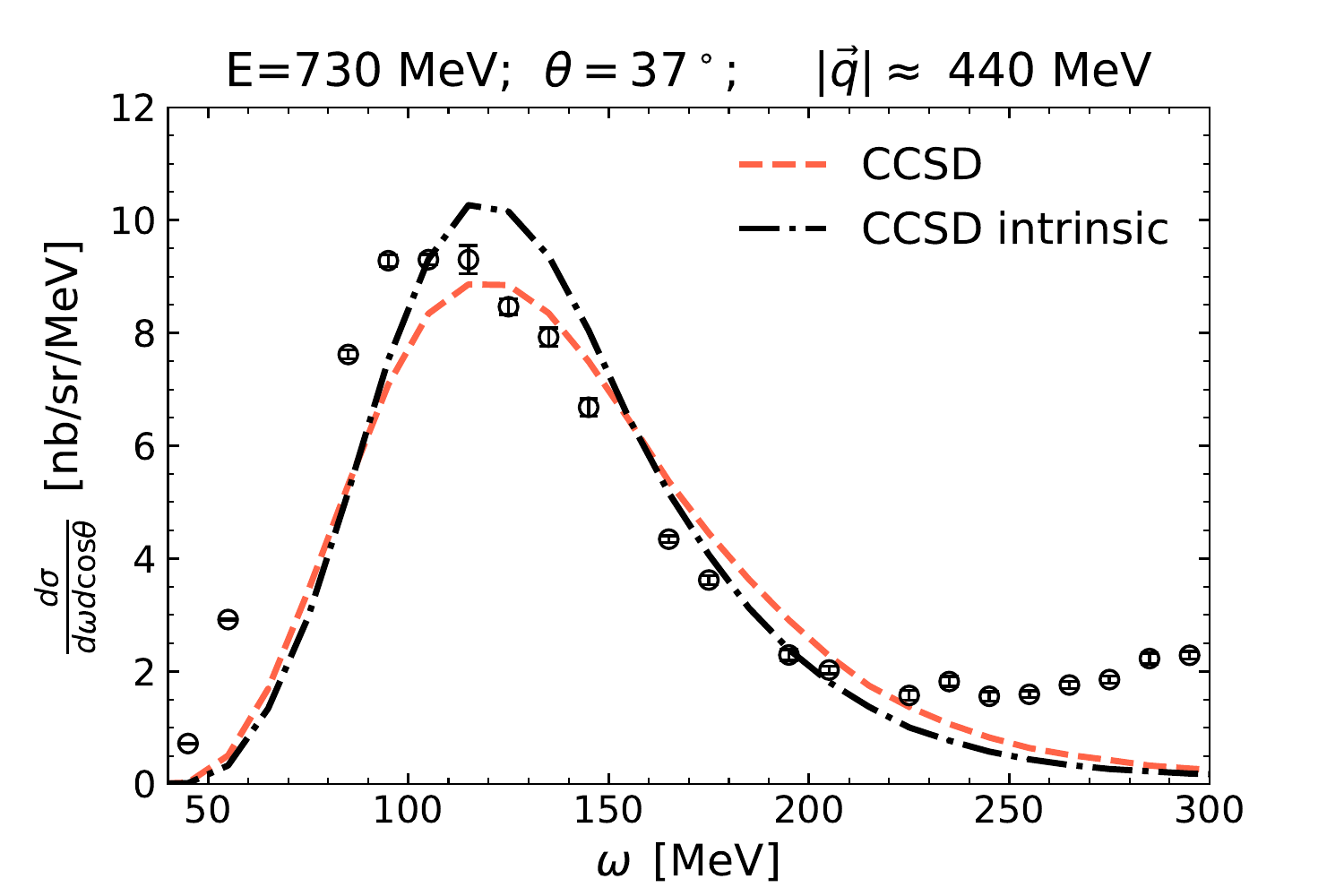}
    \includegraphics[width=0.32\textwidth]{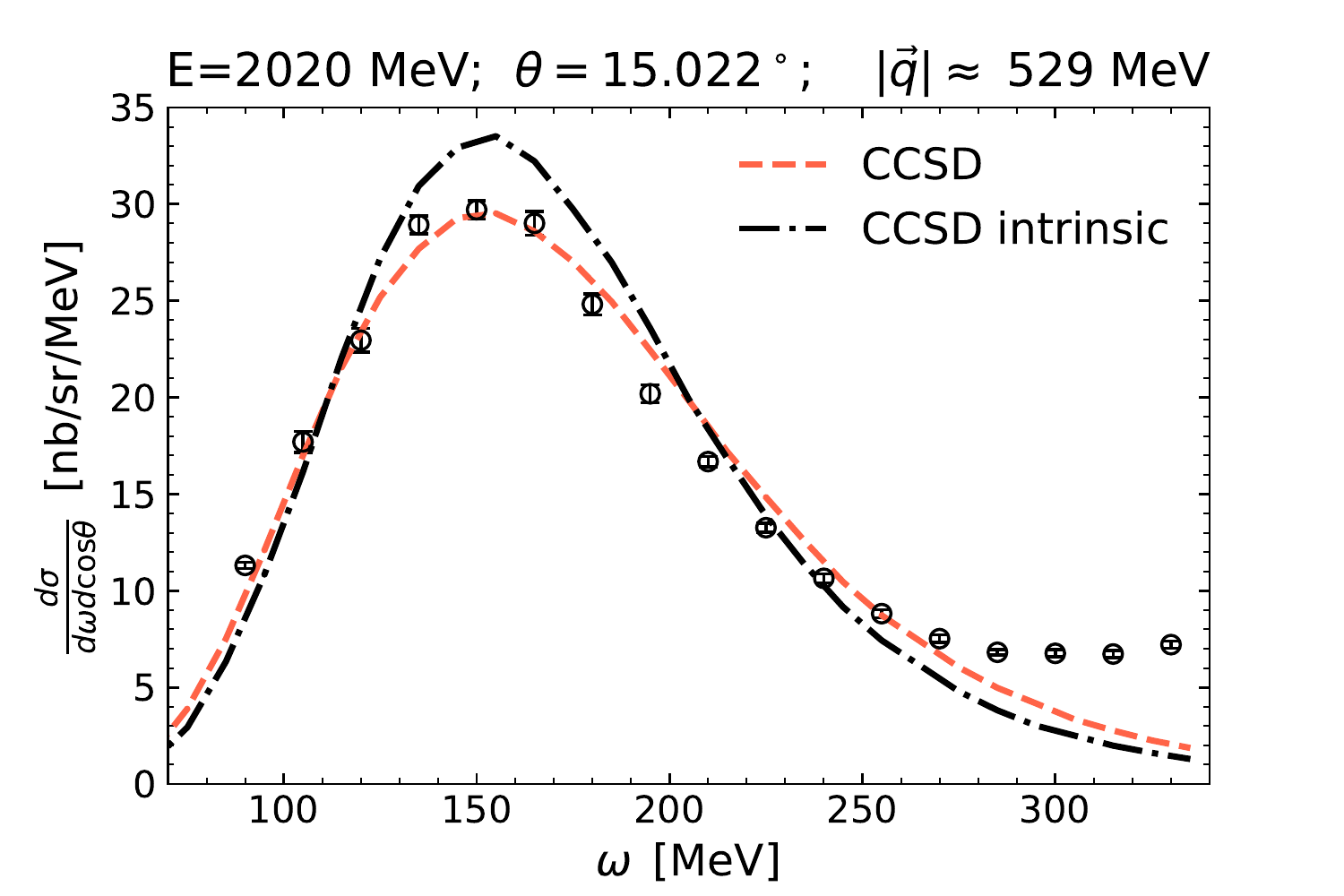}
    \includegraphics[width=0.32\textwidth]{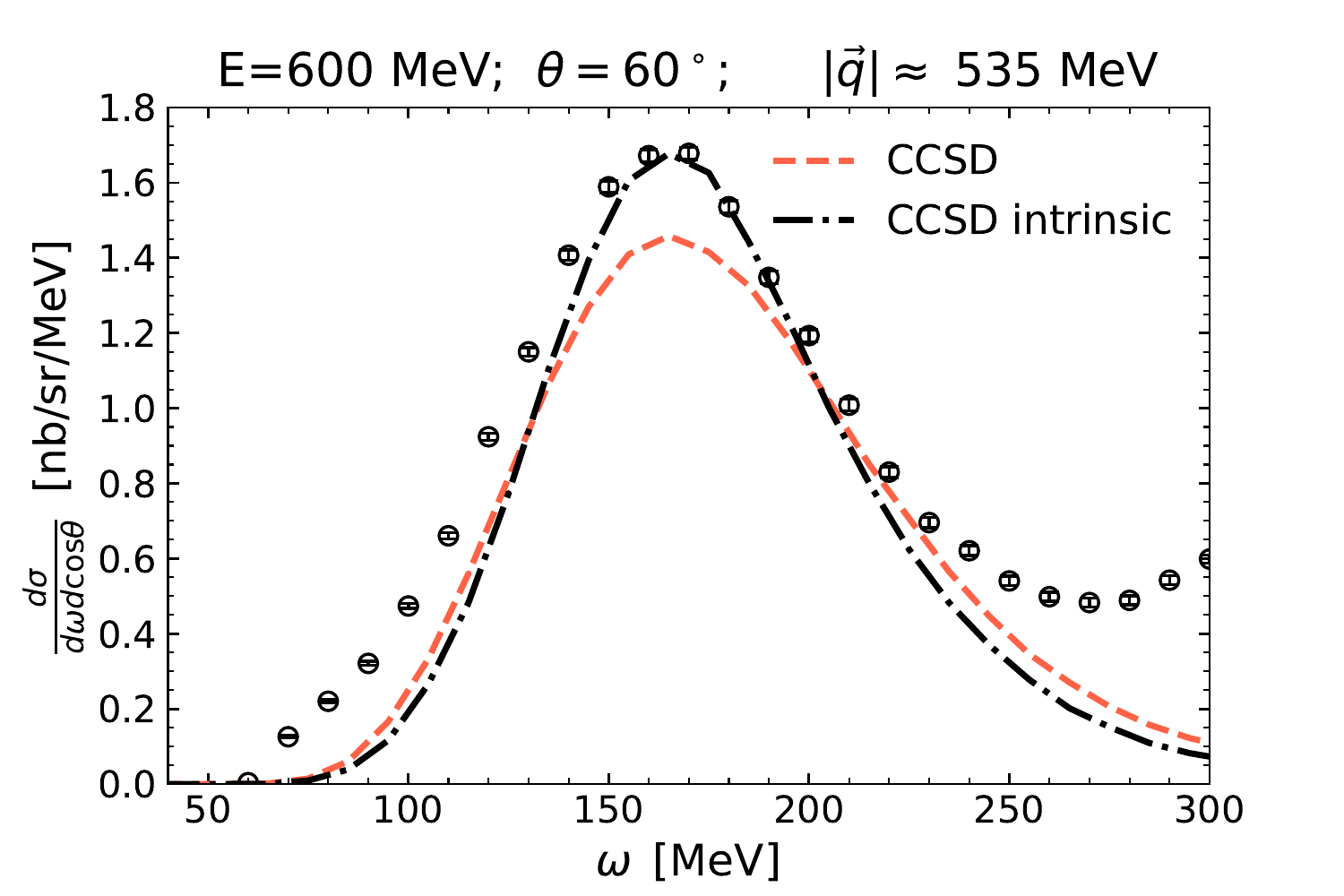}
    \includegraphics[width=0.32\textwidth]{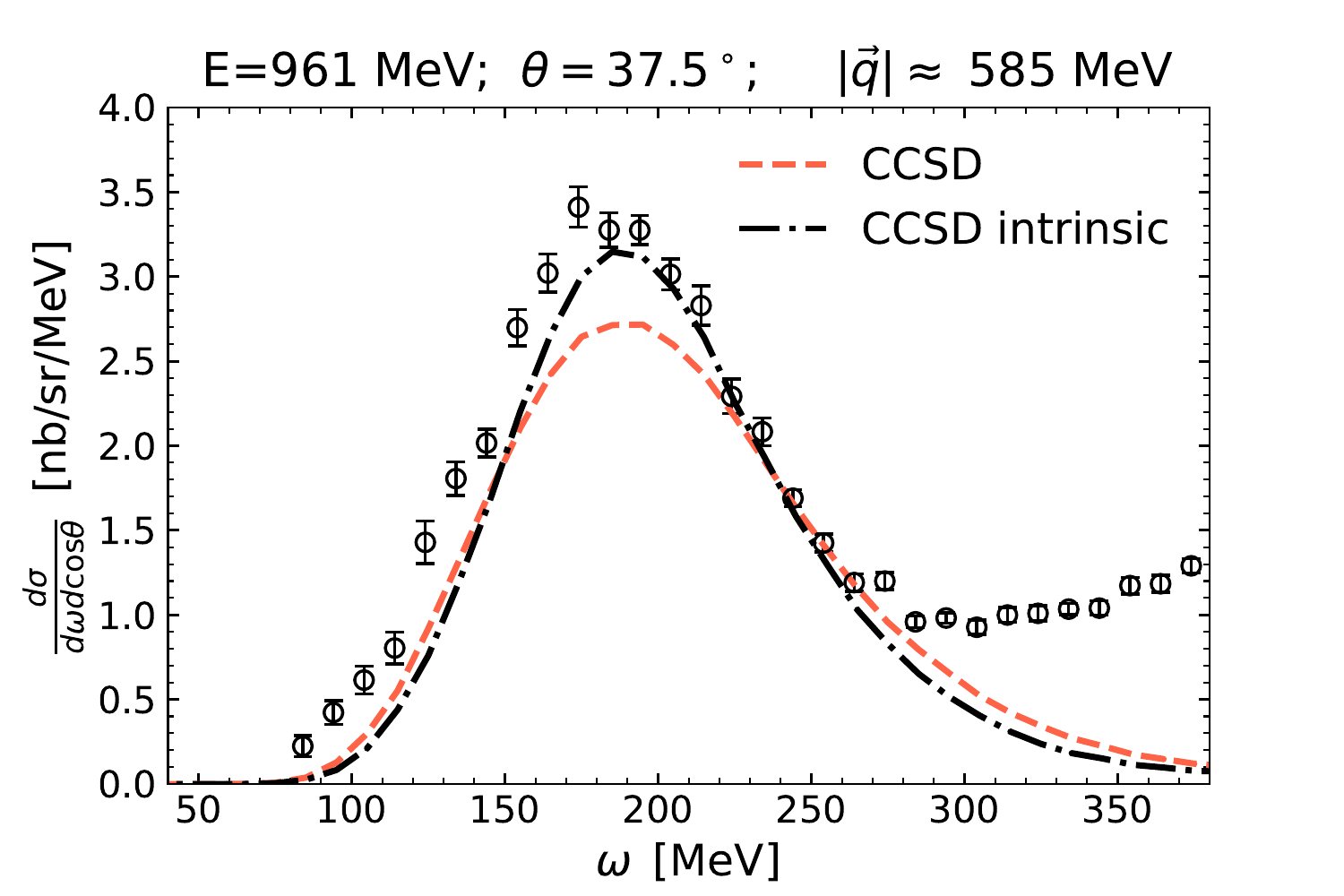}
    \includegraphics[width=0.32\textwidth]{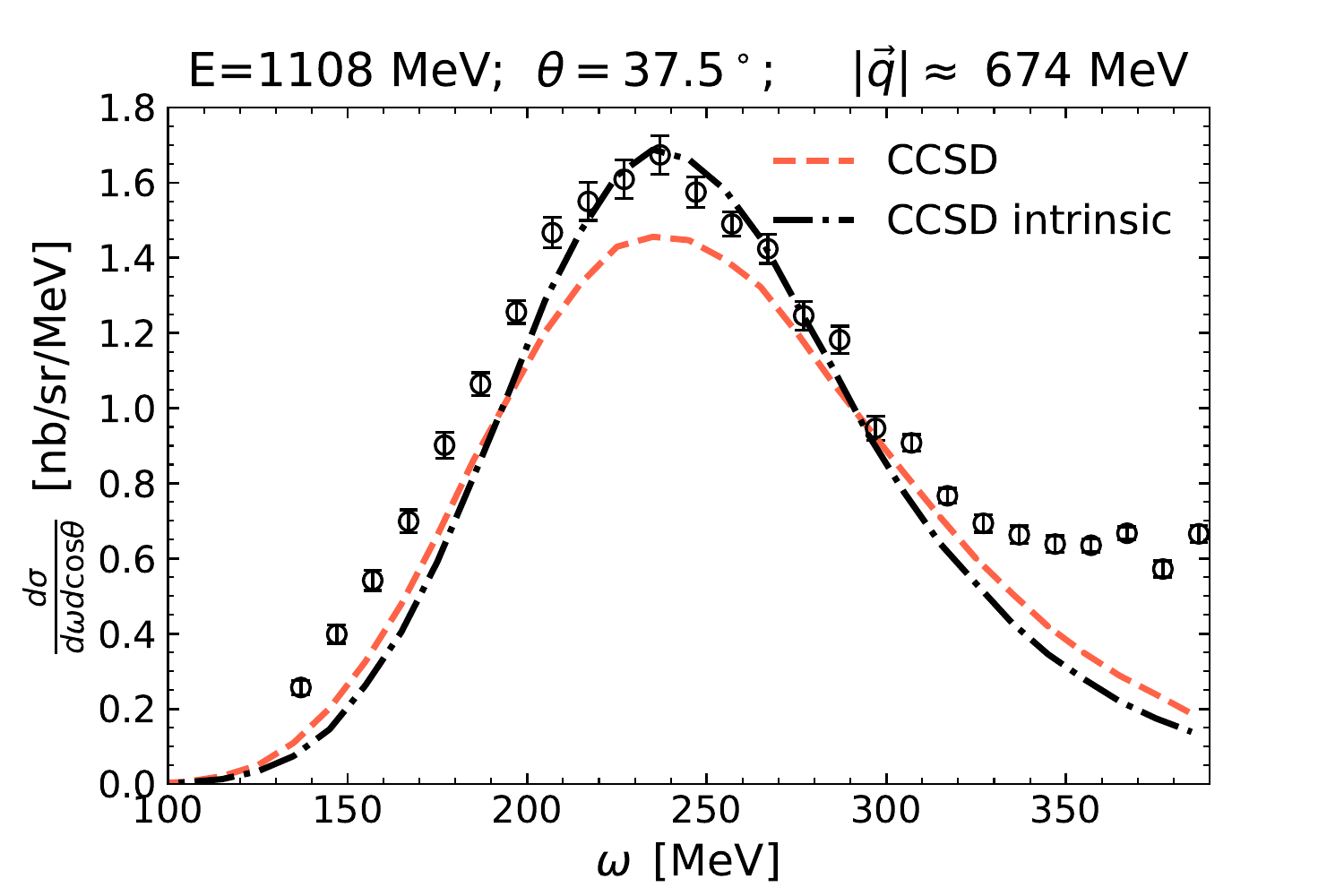}
  \caption{Electron scattering on $^4$He for different kinematics which correspond to the momentum transfers $|\mathbf{q}|\approx 270-670$~MeV. We show results obtained with CCSD before removing the center-of-mass contamination (dashed line) and afterwards, as explained in the main text (dashed-dotted line). Experimental data was taken from Ref.~\cite{O'Connell:1987ag,Zghiche:1993xg,Sealock:1989nx,Day:1993md}.}
  \label{fig:he4_electron_scattering}
\end{figure*}

We now turn our attention to the electron scattering off $^4$He.
In Fig.~\ref{fig:he4_electron_scattering}, we show results for the cross section in various kinematics, for the  spectral function before and after removal of the center-of-mass contamination. The final results, ``CCSD intrinsic'', has been obtained using the intrinsic momentum distribution. They predict more strength at the quasi-elasic peak with respect to the ``CCSD lab'' result.  This trend is consistent with the findings of Ref.~\cite{Rocco:2018vbf}. The uncertainty of $n(\mathbf{p})$ at low momenta as well as the negligible reconstruction errors coming from the ChEK method do not affect the cross-section results.  It is also interesting to notice that the impulse approximation indeed works better with increasing momentum transfer $|\mathbf{q}|$. For the values $|\mathbf{q}|\approx300-400$~MeV the  spectral function overestimates the data and predicts a shifted quasi-elastic peak.

\begin{figure}[hbt]
    \includegraphics[width=0.45\textwidth]{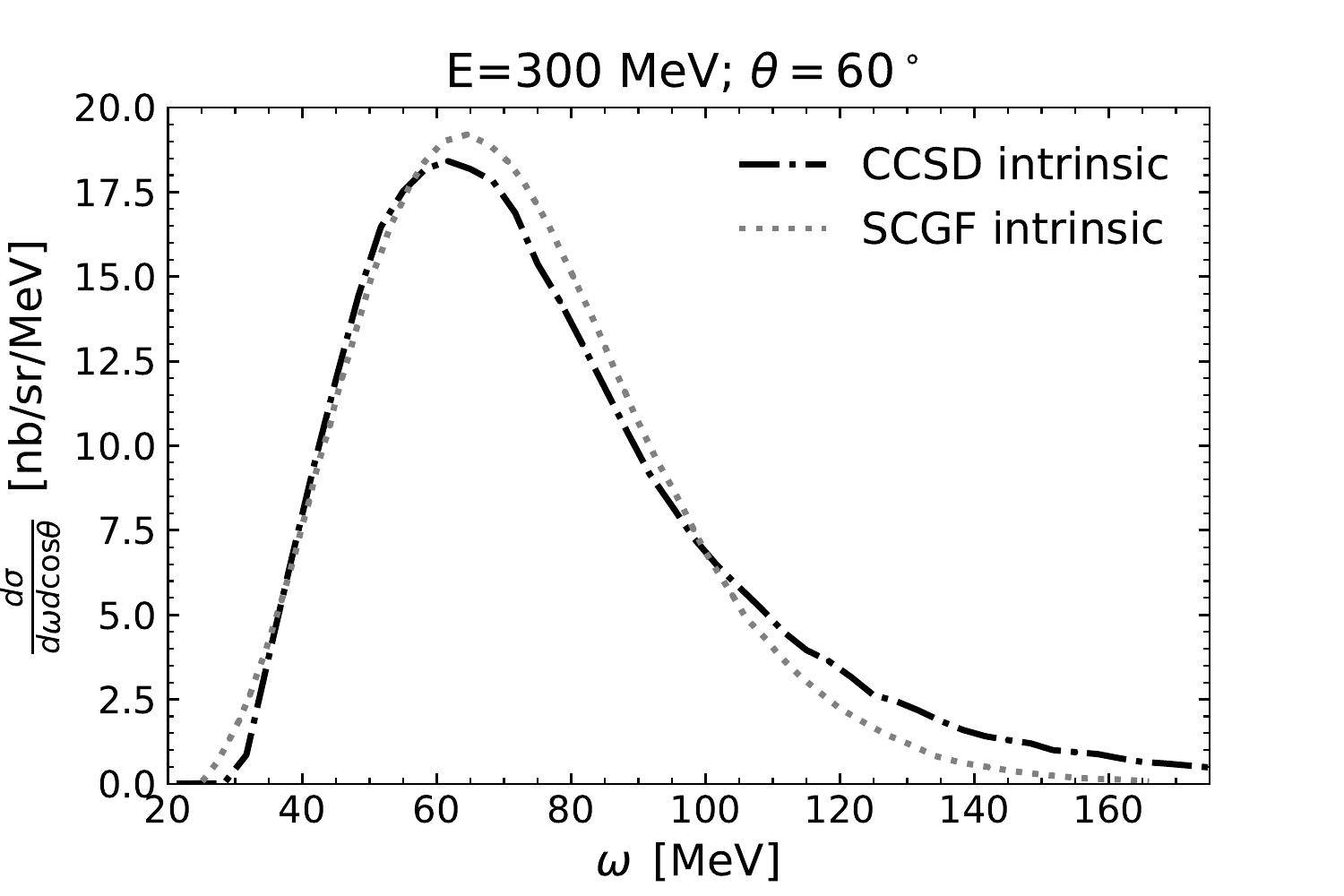}

  \caption{Comparison of CCSD results with SCFG~\cite{Rocco:2018vbf} on $^4$He after removing the center-of-mass contamination. }
  \label{fig:he4_scgf_comp}
\end{figure}

A direct comparison of our results with Ref.~\cite{Rocco:2018vbf} shows an overall good agreement. While the results before the center-of-mass removal are almost identical, the predicted cross section using the intrinsic  spectral function is slightly different, as can be seen in Fig~\ref{fig:he4_scgf_comp}. We have chosen this low momentum transfer kinematics, because the nuclear effects might play a more important role and differences between the CCSD and SCGF should be more pronounced. There are several sources of discrepancies. First, in the conservation of energy of Eq.~\eqref{eq:hadron_tens_SF} we take into account the kinetic energy of the recoiled nucleus, which for $^4$He amounts to 7-9 MeV for the Fermi momentum. Second, we use a different approach to remove the center-of-mass. Third, the spectral functions are obtained using two different many-body methods and approximations therein.

\section{Conclusion and outlook}
\label{sec:conclusion}
We have presented an ab initio calculation of the spectral functions for $^4$He  based on the coupled cluster theory combined with the 
ChEK method for the reconstruction of the spectral properties of a many-body system. Within this approach, and for a given resolution, we were able to assess the uncertainty of our calculation. For $^4$He we obtained an almost negligible error, however we expect it to be larger when we move to medium-mass nuclei.
This work  paves the way for further explorations of the ChEK method in nuclear physics. On the one hand, one can study other nuclear responses, especially where the standard inversion procedures are unable to give stable results. On the other hand, the method provides error bounds, because it does not require any ansatz about the properties of the response (e.g. the threshold energy needed in the LIT inversion). This way we can achieve a stringent control over the uncertainty bound of an observable of interest.

We compared our predictions for the electron-nucleus scattering in the quasi-elastic regime with available data and found  agreement. We were able to scan a large range of momentum transfers, and observed that the impulse approximation improves  with growing momentum. 
Our results point to  possible directions of further investigation. In view of the planned DUNE and T2HK experiments, which will benefit from reliable cross-section models,  another important step is the calculation  of spectral function for $^{16}$O and $^{40}$Ar. Both of these nuclei is within the reach of the coupled-cluster approach. 
Furthermore, for low and intermediate momentum transfers the impulse approximation picture becomes less reliable and the full inclusion of final state interactions is therefore desirable. This transition region requires more theoretical studies as well as experimental data. Recent developments within coupled-cluster theory allow us to lead a consistent analysis based on the same many-body method, nuclear dynamics and truncations, employing both LIT-CC and spectral functions. 
Lastly, our approach  does not presently account for two-body currents. Their role will also be a topic of our future work.

\begin{acknowledgements}
J.E.S. acknowledges the support of the Humboldt Foundation through a Humboldt Research Fellowship for Postdoctoral Researchers. 
This work was supported in part by the Deutsche
Forschungsgemeinschaft (DFG)
through the Cluster of Excellence ``Precision Physics, Fundamental
Interactions, and Structure of Matter" (PRISMA$^+$ EXC 2118/1) funded by the
DFG within the German Excellence Strategy (Project ID 39083149).
This material is based upon work supported by the U.S.\ Department of Energy, Office of Science, Office of Nuclear Physics under award numbers DE-FG02-96ER40963 and DE-SC0018223 (NUCLEI SciDAC-4 collaboration), and contract No. DE-AC05-00OR22725 with UT-Battelle, LLC (Oak Ridge National Laboratory). Computer time was provided by the Innovative and Novel Computational Impact on Theory and Experiment (INCITE) program and by the supercomputer Mogon at Johannes Gutenberg Universit\"at Mainz. This research used resources of the Oak Ridge Leadership Computing Facility located at Oak Ridge National Laboratory, which is supported by the Office of Science of the Department of Energy under contract No. DE-AC05-00OR22725.
\end{acknowledgements}

\appendix

\section{Removal of the center of mass}
\label{sec:appendix}
We follow Ref.~\cite{giraud2008} and consider an $A$-body system with
coordinates $\mathbf{r}_1,\ldots,\mathbf{r}_A$ and corresponding
momenta $\mathbf{p}_1,\ldots,\mathbf{p}_A$ in the laboratory
system. The center-of-mass and relative coordinates are
\ba
\mathbf{R} &=& {1\over A}\left(\mathbf{r}_1+\ldots+\mathbf{r}_A\right)\ ,\nonumber\\
\boldsymbol{\xi}_1 &=& \mathbf{r}_2 -\mathbf{r}_1\ ,\nonumber\\
\boldsymbol{\xi}_2 &=& \mathbf{r}_3 -\frac{\mathbf{r}_1+\mathbf{r}_2}{2}\ ,\nonumber\\
&\vdots&\nonumber\\
\boldsymbol{\xi}_n &=& \mathbf{r}_{n+1} -\frac{\mathbf{r}_1+\mathbf{r}_2+\ldots+\mathbf{r}_n}{n}\ ,\nonumber\\
&\vdots&\nonumber\\
\boldsymbol{\xi}_{A-1} &=& \mathbf{r}_{A} -\frac{\mathbf{r}_1+\mathbf{r}_2+\ldots+\mathbf{r}_{A-1}}{A-1} \ , 
\ea
and the corresponding canonical momenta are
\ba
\mathbf{P} &=& \mathbf{p}_1+\ldots+\mathbf{p}_A\ ,\nonumber\\
\boldsymbol{\pi}_1 &=& \frac{\mathbf{p}_2 -\mathbf{r}_1}{2}\ ,\nonumber\\
\boldsymbol{\pi}_2 &=& {2\over 3}\mathbf{p}_3 -\frac{\mathbf{p}_1+\mathbf{p}_2}{3}\ ,\nonumber\\
&\vdots&\nonumber\\
\boldsymbol{\pi}_n &=& {n\over n+1}\mathbf{p}_{n+1} -\frac{\mathbf{p}_1+\mathbf{p}_2+\ldots+\mathbf{p}_n}{n+1}\ ,\nonumber\\
&\vdots&\nonumber\\
\boldsymbol{\pi}_{A-1} &=& {A-1\over A}\mathbf{p}_{A} -\frac{\mathbf{p}_1+\mathbf{p}_2+\ldots+\mathbf{p}_{A-1}}{A} \ . 
\ea
Two comments are in order. First, the transformation between
laboratory and center-of-mass coordinates has a unit Jacobian. Second,
we see that $\boldsymbol{\xi}_n$ is the position of particle $(n+1)$
with respect to the center of mass of the previous $n$ particles,
while $\boldsymbol{\pi}_n$ is the momentum of particle $(n+1)$
relative to the average momentum of the $(n+1)$ particle system.  We
note that $\mathbf{r}_i -\mathbf{R}$ and $\mathbf{p}_i-\mathbf{P}/A$
are intrinsic positions and momenta, respectively, for any
$i=1,\ldots,A$. A particular convenient choice is $i=A$, because
\ba
\mathbf{r}_A -\mathbf{R} &=& {A-1\over A}\boldsymbol{\xi}_{A-1} \ ,
\label{specialr}
\ea
\ba
\mathbf{p}_A - {\mathbf{P}\over A} &=& \boldsymbol{\pi}_{A-1} 
\label{specialp}
\ea
can be expressed in terms of a single relative position and momentum,
respectively.

Let us now assume that the ground state
$|\Psi\rangle=|\Phi\rangle |\psi\rangle$ factorizes into a
center-of-mass state $|\Phi\rangle$ and an intrinsic state
$|\psi\rangle$. The laboratory density in position space is
\ba
\label{rhox}
\rho(\mathbf{x}) &=& A \int {\rm d}^3\mathbf{r}_1\cdots {\rm d}^3\mathbf{r}_A \delta(\mathbf{x}-\mathbf{r}_A) |\Psi(\mathbf{r}_1,\ldots,\mathbf{r}_A)|^2 \nonumber\\
&=& A \int {\rm d}^3\mathbf{r}_1\cdots {\rm d}^3\mathbf{r}_{A-1} |\Psi(\mathbf{r}_1,\ldots,\mathbf{r}_{A-1},\mathbf{x})|^2 \,.
\ea
Based on Eq.~(\ref{specialr}), the intrinsic one-body density is
\ba
\label{sigmax}
\sigma(\mathbf{x}) &=& A \int {\rm d}^3\boldsymbol{\xi}_1\cdots {\rm d}^3\boldsymbol{\xi}_{A-1}
\delta\left(\mathbf{x}-{A-1\over A}\boldsymbol{\xi}_{A-1}\right) \nonumber\\
&&\times|\psi(\boldsymbol{\xi}_1,\ldots,\boldsymbol{\xi}_{A-1})|^2 \nonumber\\
&=& A \left({A\over A-1}\right)^3\\
&&\times\int {\rm d}^3\boldsymbol{\xi}_1\cdots {\rm d}^3\boldsymbol{\xi}_{A-2}
\left|\psi\left(\boldsymbol{\xi}_1,\ldots,\boldsymbol{\xi}_{A-2},{A\over A-1}\mathbf{x}\right)\right|^2 \,.\nonumber
\ea
To establish the relation between the intrinsic and laboratory
densities we use $\delta(\mathbf{x}-\mathbf{r}_A)=
\delta(\mathbf{x}-\mathbf{R}-\boldsymbol{\xi}_{A-1}(A-1)/A)$ and
rewrite Eq.~(\ref{rhox}) as
\ba
\rho(\mathbf{x}) &=& A \int {\rm d}^3\boldsymbol{\xi}_1\cdots {\rm d}^3\boldsymbol{\xi}_{A-1} {\rm d}^3\mathbf{R}
\delta\left(\mathbf{x}-\mathbf{R}-{A-1\over A}\boldsymbol{\xi}_{A-1}\right)\nonumber\\
&&\times|\Phi(\mathbf{R})|^2 |\psi(\boldsymbol{\xi}_1,\ldots,\boldsymbol{\xi}_{A-1})|^2  \nonumber\\
&=& \int{\rm d}^3\mathbf{R} |\Phi(\mathbf{R})|^2 \sigma(\mathbf{x}-\mathbf{R}) \ .
\ea
Thus, the laboratory density is a convolution of the center-of-mass
density and the intrinsic density. In coupled-cluster and IMSRG calculations, the
center-of-mass wave function is a Gaussian (to a very good
approximation)~\cite{hagen2009a,hagen2010b,jansen2012,morris2015,parzuchowski2017}, i.e.
\be
|\Phi(\mathbf{R})|^2 = \pi^{-3/2}b^{-3} e^{-{\mathbf{R}^2\over b^2}} \ ,
\ee
and its Fourier transform is
\be
(2\pi)^{-3/2} e^{-{b^2\over 4}\mathbf{P}^2} \ .
\ee
Thus,
the intrinsic density can be obtained by dividing the Fourier
transforms of the laboratory and center-of-mass wave functions and
performing the inverse Fourier transform of that quotient.

Let us now turn to momentum space. The laboratory momentum density is
\ba
\label{rhop}
\rho(\mathbf{p}) &=& A \int {\rm d}^3\mathbf{p}_1\cdots {\rm d}^3\mathbf{p}_A \delta(\mathbf{p}-\mathbf{p}_A) |\Psi(\mathbf{p}_1,\ldots,\mathbf{p}_A)|^2 \nonumber\\
&=& A \int {\rm d}^3\mathbf{p}_1\cdots {\rm d}^3\mathbf{p}_{A-1} |\Psi(\mathbf{p}_1,\ldots,\mathbf{p}_{A-1},\mathbf{p})|^2 \,.
\ea
Based on Eq.~(\ref{specialp}), the intrinsic one-body momentum density is
\ba
\label{sigmap}
\sigma(\mathbf{p}) &=& A \int {\rm d}^3\boldsymbol{\pi}_1\cdots {\rm d}^3\boldsymbol{\pi}_{A-1}
\delta\left(\mathbf{p}-\boldsymbol{\xi}_{A-1}\right) \nonumber\\
&&\times|\psi(\boldsymbol{\pi}_1,\ldots,\boldsymbol{\pi}_{A-1})|^2 \nonumber\\
&=& A \int {\rm d}^3\boldsymbol{\pi}_1\cdots {\rm d}^3\boldsymbol{\pi}_{A-2}
\left|\psi\left(\boldsymbol{\pi}_1,\ldots,\boldsymbol{\pi}_{A-2},\mathbf{p}\right)\right|^2 \ .\nonumber\\
\ea
To establish the relation between the intrinsic and laboratory momentum densities we use
$\delta(\mathbf{p}-\mathbf{p}_A)=\delta(\mathbf{p}-\boldsymbol{\pi}_{A-1}-\mathbf{P}/A)$ and
rewrite Eq.~(\ref{rhop}) as
\ba
\rho(\mathbf{p}) &=& A \int {\rm d}^3\boldsymbol{\pi}_1\cdots {\rm d}^3\boldsymbol{\pi}_{A-1} {\rm d}^3\mathbf{P}
\delta\left(\mathbf{p}-\boldsymbol{\pi}_{A-1}-{\mathbf{P}\over A}\right)\nonumber\\
&&\times|\Phi(\mathbf{P})|^2 |\psi(\boldsymbol{\pi}_1,\ldots,\boldsymbol{\pi}_{A-1})|^2  \nonumber\\
&=& \int{\rm d}^3\mathbf{P} |\Phi(\mathbf{P})|^2 \sigma\left(\mathbf{p}-{\mathbf{P}\over A}\right) \ .
\ea
The substitition $\mathbf{K}=\mathbf{P}/A$ then yields 
\ba
\rho(\mathbf{p}) = A^3\int{\rm d}^3\mathbf{K} |\Phi(A\mathbf{K})|^2 \sigma\left(\mathbf{p}-\mathbf{K}\right) \ .
\ea
Thus, the laboratory density is a convolution of the center-of-mass
density (at $A$ times its argument) and the intrinsic density. Again,
the center-of-mass wave function is to a good approximation a
Gaussian in momentum space, and the deconvolution can be performed. We have
\be
\label{gaussP}
A^3|\Phi(A\mathbf{P})|^2 = \pi^{-3/2}b^{3}A^3 e^{-b^2A^2\mathbf{P}^2} \ ,
\ee
and its Fourier transform is
\be
(2\pi)^{-3/2} e^{-{1\over 4}{\mathbf{R}^2\over A^2b^2}} \ .
\ee
In the Gaussians, we employed the oscillator length
\be
b = {\sqrt{\hbar\over M\tilde{\omega}}} \ .
\ee
Here, $M=Am$ is the total mass in terms of the nucleon mass $m$, while
$\tilde{\omega}$ is the frequency of the Gaussian; this parameter is
independent of the oscillator basis, and we have
$\hbar\tilde{\omega}\approx 24$~MeV for a light nucleus such as $^4$He
and $\hbar\tilde{\omega}\approx 16-20$~MeV for
$^{16}$O\cite{hagen2009a,hagen2010b}. Thus $b\approx 0.7A^{-1/2}$~fm,
and the Gaussian~(\ref{gaussP}) approximates a $\delta$-function for
$A\gg 1$. We therefore expect that the difference between the
laboratory and intrinsic density in momentum space becomes
exponentially small for $A^{1/2}\gg 1$.

\bibliography{biblio}

\end{document}